\documentclass{article}
\usepackage{amsmath,amsthm,amsfonts,amscd,eucal}

\numberwithin{equation}{section}

\def\bn{{\mathbb N}}
\def\br{{\mathbb R}}

\def\itm#1{\item{$(#1)$}}

\DeclareMathOperator{\Var}{Var}

\newtheorem{theorem}{Theorem}[section]
\newtheorem{proposition}[theorem]{Proposition}

\newtheorem{corollary}[theorem]{Corollary}
\theoremstyle{definition}
\newtheorem{definition}[theorem]{Definition}
\newtheorem{example}[theorem]{Example}
\newtheorem{remark}[theorem]{Remark}

\makeindex

\begin{document}

\title{
{Inequalities for quantum Fisher information}
}

\author{
Paolo Gibilisco\footnote{Dipartimento SEFEMEQ and Centro V.Volterra, Facolt\`a di
Economia, Universit\`a di Roma ``Tor Vergata", Via Columbia 2, 00133
Rome, Italy.  Email:  gibilisco@volterra.uniroma2.it --
URL: http://www.economia.uniroma2.it/sefemeq/professori/gibilisco},
Daniele Imparato\footnote{Dipartimento di Matematica, Politecnico di Torino, 
Corso Duca degli Abruzzi 24, 10129 Turin, Italy. 
Email: daniele.imparato@polito.it}
\
and
Tommaso Isola\footnote{Dipartimento di Matematica, 
Universit\`a di Roma ``Tor Vergata", 
Via della Ricerca Scientifica, 00133 Rome, Italy.
Email: isola@mat.uniroma2.it
URL: http://www.mat.uniroma2.it/$\sim$isola}
}

\maketitle

\begin{abstract}

In the paper \cite{Luo:2003a} Luo proved
an inequality relating the Wigner-Yanase information and the $SLD$-information. In this paper we prove that Luo's inequality is a particular case of a general inequality which holds for any regular quantum Fisher information. Moreover we show that this general inequality is a consequence of the Kubo-Ando inequality that states that any matrix mean is bigger than the harmonic mean and smaller than the arithmetic mean.



\end{abstract}

\section{Introduction}

Fisher information appeared for the first time in \cite{Fisher:1925}. From that seminal work the use of Fisher information spread out, not only in statistics, but also in other mathematical fields,  and in a number of applied sciences \cite{Frieden:2004}. Several quantum versions of Fisher information have been studied. Among the first examples one has the Wigner-Yanase information (see \cite{WignerYanase:1963}
or \cite{GibiliscoIsola:2001}\cite{GibiliscoIsola:2003}\cite{GibiliscoIsola:2004}\cite{GibiliscoIsola:2005} for a recent treatment) and the $SLD$-information (see \cite{Bures:1969}\cite{Uhlmann:1992}\cite{Helstrom:1969}) that are defined as follows.
As usual $[\cdot,\cdot]$ denotes the commutator. Let $\rho$ be a density matrix and let $A$ be a self-adjoint matrix. Let $L$ be the solution of the operator equation $(L\rho+\rho L)=2i[\rho,A].$
Define the Wigner-Yanase and the $SLD$-information as
\begin{equation}\label{eq1.0}
I_{\rho}^{WY}(A):=-\frac{1}{2}{\rm Tr }([\rho^{\frac{1}{2}},A]^2), \qquad \qquad I^{SLD}_{\rho}(A):=\frac{1}{4}{\rm Tr}(\rho L^2).
\end{equation}
In the paper \cite{Luo:2003a} Luo proved the following three results.

i) If $\rho(t):=e^{-itA}\rho e^{itA}$, the functions of $t$ given by $I_{\rho(t)}^{WY}(A),I_{\rho(t)}^{SLD}(A)$ are constant (this is Theorem 1 in \cite{Luo:2003a}).

ii) The following inequality is true (this is Theorem 2 in \cite{Luo:2003a}):
\begin{equation}\label{eq1.1}
I_{\rho}^{WY}(A) \leq I_{\rho}^{SLD}(A)\leq 2I_{\rho}^{WY}(A). 
\end{equation}

iii) The constant 2 is optimal in the inequality (\ref{eq1.1}). Namely, if $1 \leq k <2$, the inequality

$$
I_{\rho}^{SLD}(A)\leq kI_{\rho}^{WY}(A)
$$
is false, and a counterexample can be found in the elementary $ 2 \times 2$ case (this is the final Example in \cite{Luo:2003a}).

A full quantum theory for Fisher information was established only  a few years ago by  Petz in his classification theorem \cite{Petz:1996}. It is worth to note that the Petz theorem rests on two fundamental breakthroughs due to Rao and Chentsov. Rao observed that Fisher information should be seen as a Riemannian metric on statistical models \cite{Rao:1945}. Chentsov characterized Fisher information as the unique (in the appropriate setting) Riemannian metric contracting under coarse graining 
\cite{Chentsov:1982}.

Starting from this idea, Petz defined the quantum Fisher informations (QFI) as Riemannian metrics (on the state manifold) contracting under coarse graining. He was able to prove that QFI are parametrized by functions $f \in {\cal F}_{op}$, where ${\cal F}_{op}$ is the set of symmetric normalized operator monotone functions. The regular elements of ${\cal F}_{op}$ are those for which $f(0)>0$. The corresponding QFI is said regular too. For regular QFI one can define the metric adjusted skew information (or $f$-information) as 
$$
I^f_{\rho}(A):= \frac{f(0)}{2}||i[\rho,A]||_{\rho,f}
$$
(see \cite{Hansen:2006b} \cite{GibiliscoIsola:2007}). The $WY$ and $SLD$ informations, defined in (\ref{eq1.0}), are particular cases of the above definition.

In this paper we show that the three results proved by Luo are particular cases of the following general results.

i') Set $\rho_H(t):=e^{-itH}\rho e^{itH}$. If $[A,H]=0$ then the function $I^f_{\rho_H(t)}(A)$ is constant. Since quantum Fisher informations contract under coarse graining they are unitary covariant and this is the crucial ingredient of the proof. This result was stated by Hansen  in \cite{Hansen:2006b} and we provide here a detailed proof.

ii') The inequality (\ref{eq1.1}) is a particular case of the following inequality
\begin{equation}\label{eq1.2}
I_{\rho}^{f}(A) \leq I_{\rho}^{SLD}(A)\leq \frac{1}{2f(0)}I_{\rho}^{f}(A), 
\end{equation}
which is true for any (regular) quantum Fisher information. Inequality (\ref{eq1.2}) is a consequence of the Kubo-Ando inequality
$$
2(A^{-1}+B^{-1})^{-1} \leq m(A,B)\leq \frac{A+B}{2}
$$
that states that any matrix mean is bigger then harmonic mean and smaller then arithmetic mean.

iii') The constant $\frac{1}{2f(0)}$ is optimal in inequality (\ref{eq1.2}). Namely, if $1 \leq k < \frac{1}{2f(0)}$, the inequality

$$
I_{\rho}^{SLD}(A)\leq kI_{\rho}^{f}(A)
$$
is false and a counterexample can be found in the elementary $ 2 \times 2$ case. 

Let us observe that in the papers \cite{Luo:2000}  \cite{Luo:2003b} Luo proved also another inequality for the $WY$ and $SLD$ information, namely
\begin{equation}\label{eq1.3}
I_{\rho}^{WY}(A)\leq {\rm Var}_{\rho}(A), \qquad \qquad I_{\rho}^{SLD}(A)\leq {\rm Var}_{\rho}(A). 
\end{equation}

From inequalities (\ref{eq1.2}) and (\ref{eq1.3}) one immediately obtains that also this result is completely general, namely
$$
I_{\rho}^{f}(A)\leq {\rm Var}_{\rho}(A), 
$$
a result recently proved by Hansen in \cite{Hansen:2006b} and with a different approach by ourselves in \cite{GibiliscoIsola:2007}.

\section{Operator monotone functions, matrix means and quantum Fisher information}

Let $M_n:=M_n(\mathbb{C})$ (resp. $M_{n,sa}:=M_n(\mathbb{C})_{sa}$) be
the set of all $n \times n$ complex matrices (resp.  all $n \times n$
self-adjoint matrices).  We shall denote general matrices by $X,Y,...$
while letters $A,B,...$ will be used for self-adjoint matrices (the
Hilbert-Schmidt scalar product is denoted by $\langle A,B \rangle={\rm
Tr}(A^*B)$).  The adjoint of a matrix $X$ is denoted by $X^{\dag}$
while the adjoint of a superoperator $T:(M_n,\langle \cdot,\cdot
\rangle) \to (M_n ,\langle \cdot,\cdot \rangle)$ is denoted by $T^*$. 
Let ${\cal D}_n$ be the set of strictly positive elements of $M_n$
while ${\cal D}_n^1 \subset {\cal D}_n$ is the set of strictly
positive density matrices, namely
$
{\cal D}_n^1=\{\rho \in M_n \vert {\rm Tr} \rho=1, \, \rho>0 \}
$.
If it is not specified from now on we treat the case of faithful states, namely $\rho>0$. 

\begin{definition}
Suppose that $\rho \in {\cal D}_n^1$ is fixed.  Define $X_0:=X-{\rm
Tr}(\rho X) I$.
\end{definition}

\begin{definition}
For $A,B \in M_{n,sa}$ and $\rho \in {\cal D}_n^1$ define covariance
and variance as
$$
{\rm Cov}_{\rho}(A,B):={\rm Tr}(\rho A B)-{\rm Tr}(\rho A)\cdot{\rm
Tr}(\rho B)= {\rm Tr}(\rho A_0 B_0)
$$
$$
{\rm Var}_{\rho}(A):={\rm Tr}(\rho A^2)-{\rm Tr}(\rho A)^2 = {\rm
Tr}(\rho A^2_0).
$$
\end{definition}

Let ${\mathbb R}^+:=(0,\infty)$. A function $f:{\mathbb R}^+\to \mathbb{R}$ is said
{\it operator monotone (increasing)} if, for any $n\in \bn$, any $A$, $B\in M_n$
such that $0\leq A\leq B$, the inequalities $0\leq f(A)\leq f(B)$
hold.  An operator monotone function is said {\it symmetric} if
$f(x)=xf(x^{-1})$ and {\it normalized} if $f(1)=1$.

\begin{definition}
${\cal F}_{op}$ is the class of functions $f: {\mathbb R}^+
\to{\mathbb R}^+$ such that

\itm{i'} $f(1)=1$,

\itm{ii'} $tf(t^{-1})=f(t)$,

\itm{iii'} $f$ is operator monotone.
\end{definition}

\begin{example}
Two  important elements of ${\cal F}_{op}$ are
$$
f_{WY}(x):=\left(\frac{1+\sqrt{x}}{2}\right)^2, \qquad \qquad f_{SLD}(x)=\frac{1+x}{2}.
$$
\end{example}

 We now report Kubo-Ando theory of matrix means (see
\cite{KuboAndo:1979/80}) as exposed in \cite{PetzTemesi:2005}.

\begin{definition}
 A {\sl mean} for pairs of positive matrices is a function
$m:{\cal D}_n \times {\cal D}_n \to {\cal D}_n$ such that

(i) $m(A,A)=A$,

(ii) $m(A,B)=m(B,A)$,

(iii) $A <B  \quad \Longrightarrow \quad A<m(A,B)<B$,

(vi) $A<A', \quad B<B' \quad \Longrightarrow \quad m(A,B)<m(A',B')$,

(v) $m$ is continuous,

(vi) $Cm(A,B)C^* \leq m(CAC^*,CBC^*)$, for every $ C \in M_n$.
\end{definition}

Property $(vi)$ is known as the transformer inequality. We denote by $\displaystyle {\cal M}_{op}$ the set of matrix means. The fundamental result, due to Kubo and Ando, is the following

\begin{theorem}
There exists a bijection between ${\cal M}_{op}$ and ${\cal F}_{op}$ given by
the formula
$$
m_f(A,B):= A^{\frac{1}{2}}f(A^{-\frac{1}{2}} B
A^{-\frac{1}{2}})A^{\frac{1}{2}}.
$$
\end{theorem}

When $A$ and $B$ commute (for example if $A=x, B=y$ are positive numbers) we have that
$$
m_f(A,B):= A \cdot f( B  A^{-1}).
$$

\begin{example}
The  arithmetic, geometric and harmonic (matrix) means are given respectively by
$$
m_{\cal A}(A,B):=A \nabla B:=\frac{1}{2}(A+B),
$$
$$
m_{\cal G}(A,B):=A\# B:=A^{\frac{1}{2}}(A^{-\frac{1}{2}} B
A^{-\frac{1}{2}})^{\frac{1}{2}}A^{\frac{1}{2}},
$$
$$
m_{\cal H}(A,B):=A{\rm !}B:=2(A^{-1}+B^{-1})^{-1}.
$$
\end{example}

The convex combination of two means is still a mean (see \cite{KuboAndo:1979/80}). 
Kubo and Ando  \cite{KuboAndo:1979/80} proved that, among matrix means, arithmetic is the largest while harmonic is the
smallest.

\begin{corollary} \label{basic}
For any $f \in {\cal F}_{op}$ and for any $x,y>0$ one has 
$$
f_{RLD}(x):=\frac{2x}{1+x}\leq f(x) \leq \frac{1+x}{2}, 
$$
$$
\frac{2xy}{x+y}\leq m_f(x,y) \leq \frac{x+y}{2}. 
$$
\end{corollary}

In what follows if ${\cal N}$ is a differential manifold we denote by
$T_{\rho} \cal N$ the tangent space to $\cal N$ at the point $\rho \in
{\cal N}$.  Recall that there exists a natural identification
 of $T_{\rho}{\cal D}^1_n$ with the space of self-adjoint traceless
 matrices; namely, for any $\rho \in {\cal D}^1_n $
$$
T_{\rho}{\cal D}^1_n =\{A \in M_n|A=A^* \, , \, \hbox{Tr}(A)=0 \}.
$$

 A Markov
morphism is a completely positive and trace preserving operator $T:
M_n \to M_m$.
A monotone metric (also said a quantum Fisher infromation)  is a family of Riemannian metrics $g=\{g^n\}$
 on $\{{\cal D}^1_n\}$, $n \in \mathbb{N}$, such that
 $$
 g^m_{T(\rho)}(TX,TX) \leq g^n_{\rho}(X,X)
 $$
 holds for every Markov morphism $T:M_n \to M_m$, for every $\rho \in
 {\cal D}^1_n$ and for every $X \in T_\rho {\cal D}^1_n$.
Usually monotone metrics are normalized in such a way that 
$[A,\rho]=0$ implies $g_{f,\rho} (A,A)={\rm Tr}({\rho}^{-1}A^2)$.

Define $L_{\rho}(A):= \rho A$, and $R_{\rho}(A):= A\rho$, and observe
 that they are commuting self-adjoint superoperators on $M_{n,sa}$.  
Now we can state the fundamental theorems about monotone metrics.

\begin{theorem} (see \cite{Petz:1996}) 
	
    There exists a bijective correspondence between monotone metrics (quantum Fisher informations) 
    on ${\cal D}^1_n$ and normalized symmetric operator monotone
    functions $f\in {\cal F}_{op}$.  This correspondence is given by
    the formula
    $$
   \langle A,B \rangle_{\rho,f}:={\rm Tr}(A\cdot
    m_f(L_{\rho},R_{\rho})^{-1}(B)).
    $$
\end{theorem}
We set $||A||^2_{\rho,f} := \langle A,A \rangle_{\rho,f}$.

\begin{proposition} \label{bounds}
$$
||A||_{\rho,f_{SLD}} \leq ||A||_{\rho,f} \leq ||A||_{\rho,f_{RLD}}.
$$
\end{proposition}
\begin{proof}
Immediate consequence of Corollary \ref{basic}.
\end{proof}

\begin{proposition}\label{unitary} (See \cite{Petz:1996} pag. 83) Monotone metrics are unitarily covariant, namely if $U$ is unitary then
$$
||U^*AU||^2_{U^*\rho U, f}=||A||^2_{\rho,f}.
$$
\end{proposition}

\section{The function $\tilde f$ and the $f$-information}

For $f \in {\cal F}_{op}$ define $f(0):=\lim_{x\to 0} f(x)$.
The condition $f(0)\not=0$ is relevant because it is a necessary and
sufficient condition for the existence of the so-called radial
extension of a monotone metric to pure states (see
\cite{PetzSudar:1996}). 
Following \cite{Hansen:2006b} we say that a function $f \in {\cal
F}_{op}$ is {\sl regular} iff $f(0) \not= 0$.  The corresponding
operator mean, associated QFI, etc.  are said regular
too.

\begin{definition}
$$
{\cal F}_{op}^{\, r}:=\{f\in {\cal F}_{op}| \quad f(0) \not= 0 \},  \quad
{\cal F}_{op}^{\, n}:=\{f\in {\cal F}_{op}| \quad f(0) = 0 \}.
$$
\end{definition}

Trivially one has ${\cal F}_{op}={\cal F}_{op}^{\, r}\dot{\cup}{\cal F}_{op}^{\, n}$.

\begin{definition}
For $f \in {\cal F}_{op}^{\, r}$ and $x>0$ set
$$
\tilde{f}(x):=\frac{1}{2}\left[ (x+1)-(x-1)^2 \frac{f(0)}{f(x)}
\right].  
$$
\end{definition}

\begin{example} \label{serve}
$$
{\tilde f}_{WY}(x)=\sqrt{x},\qquad \qquad {\tilde f}_{SLD}(x)=\frac{2x}{1+x}.
$$
\end{example}

Observe \cite{GibiliscoIsola:2007} that $
f \in {\cal F}_{op}^{\, r}$ implies  ${\tilde f} \in {\cal
F}_{op}^{\, n}$.

A self-adjoint operator $A$ determines the evolution of the state $\rho$ by the formula
$
\rho_A(t):=e^{-iAt}\rho e^{iAt}$.
The evolution satisfies the equation
$
\dot{\rho}_A(t)=i[\rho_A(t),A]
$.
We set
$$
\dot{\rho}_A:=\dot{\rho}_A(0)=i[\rho,A].
$$
Observe that $L:=2(L_{\rho}+R_{\rho})^{-1}(i[\rho,A])$  can be seen as a quantum analogue of the symmetric logarithmic derivative (see \cite{Luo:2003a}).

\begin{definition}
$$
I_{\rho}^{WY}(A):=-\frac{1}{2}{\rm Tr}([\rho^{\frac{1}{2}},A]^2), \qquad \qquad I^{SLD}_{\rho}(A):= \frac14 {\rm Tr}\bigl(\rho L^2\bigr).
$$
\end{definition}

\begin{proposition}
$$
I_{\rho}^{WY}(A)=\frac{f_{WY}(0)}{2}|| \dot{\rho}_A ||^2_{\rho,f_{WY}}, \qquad \qquad I_{\rho}^{SLD}(A)=\frac{f_{SLD}(0)}{2}|| \dot{\rho}_A ||^2_{\rho,f_{SLD}}.
$$
\end{proposition}
\begin{proof}
For the first equality see \cite{HasegawaPetz:1997} or \cite{GibiliscoIsola:2001}\cite{Hansen:2006b}. For the second equality remember that
$
f_{SLD}(x):=\frac{1+x}{2}
$.

Therefore  one has
\begin{align*}
I_{\rho}^{SLD}(A)
&={\rm Tr}\bigl( \rho (L_{\rho}+R_{\rho})^{-1}(i[\rho,A])(L_{\rho}+R_{\rho})^{-1}(i[\rho,A]) \bigr)\\
&=\frac{1}{2} {\rm Tr}\bigl((L_{\rho}+R_{\rho}) (L_{\rho}+R_{\rho})^{-1}(\dot{\rho}_A)(L_{\rho}+R_{\rho})^{-1}(\dot{\rho}_A) \bigr)\\
&=\frac{1}{4}{\rm Tr}(2 (L_{\rho}+R_{\rho})^{-1}(\dot{\rho}_A)(\dot{\rho}_A))\\
&=\frac{f_{SLD}(0)}{2} {\rm Tr}(m_{SLD}(L_{\rho},R_{\rho})^{-1}(\dot{\rho}_A)(\dot{\rho}_A))\\
&=\frac{f_{SLD}(0)}{2}|| \dot{\rho}_A ||^2_{\rho,f_{SLD}}.
\end{align*}
\end{proof}

\begin{definition} 
For $f\in {\cal F}^r_{op}$ the metric adjusted skew information (or $f$-information) is defined as
$$
I_{\rho}^{f}(A):=\frac{f(0)}{2}|| \dot{\rho}_A ||^2_{\rho,f}.
$$
\end{definition}

Of course, if $\rho$ and $A$ commute then $I_{\rho}^{f}(A)=0$.
In what follows the following definition is very important.

\begin{definition}
$$
{\cal C}^f_{\rho}(A_0):={\rm Tr}(m_{f}(L_{\rho},R_{\rho})(A_0) \cdot A_0).
$$
\end{definition}

Observe \cite{GibiliscoIsola:2007} that 
$
I_{\rho}^f(A)={\rm Var}_{\rho}(A)-{\cal C}^{\tilde f}_{\rho}(A_0)
$.
Note that this formula allows us to consider the $f$-information also for not faithful states.

\begin{definition} 
For any state (faithful or not faithful) and for $f$ regular define:
$$
I_{\rho}^f(A):={\rm Var}_{\rho}(A)-{\cal C}^{\tilde f}_{\rho}(A_0).
$$
\end{definition}

\begin{proposition}\label{boh} (See \cite{GibiliscoIsola:2007}).
\end{proposition}
$$
\begin{array}{rcl}
g \leq f \qquad \qquad& \Longrightarrow &\qquad \qquad 0 \leq {\cal C}^g_{\rho}(A_0) \leq {\cal C}^f_{\rho}(A_0) \\[12pt]
\rho \hbox{ pure }\qquad \qquad &\Longrightarrow &\qquad \qquad {\cal C}^g_{\rho}(A_0)=0. \\
\end{array}
$$

We have immediately the following result.
\begin{proposition} 
$$
I_{\rho}^f(A) \leq {\rm Var}_{\rho}(A)
$$
with equality on pure states.
\end{proposition}

Luo (see \cite{Luo:2005a}) suggested that if one considers the variance as a measure of ``uncertainty" of an observable $A$ in the state $\rho$ then the  equality 
$$
{\rm Var}_{\rho}(A)=I_{\rho}^f(A)+{\cal C}^{\tilde f}_{\rho}(A_0)
$$
splits the variance in a ``quantum" part ($I_{\rho}^f(A)$) plus a ``classical" part (${\cal C}^{\tilde f}_{\rho}(A_0)$).

\section{The main results}

Theorem 1 in \cite{Luo:2003a} is a particular case of the following result (that was stated by Hansen in \cite{Hansen:2006b}). 

\begin{theorem}  If $[A,H]=0$ then
$I^f_{\rho_H(t)}(A)=I^f_{\rho}(A)$, for all $t\in\br$.
\end{theorem}
\begin{proof}
\end{proof}
Set $U_t:=e^{itH}$ then
$$
\rho_H(t):=e^{-itH} \rho e^{itH}=U_t^* \rho U_t.
$$
Since $[A,U_t]=0$ we have (using Proposition \ref{unitary})
\begin{align*}
I^f_{\rho_H(t)}(A)
&=\frac{f(0)}{2}||i[\rho_H(t),A]||^2_{\rho_H(t),f}=\frac{f(0)}{2}||i[U_t^* \rho U_t,A]||^2_{U_t^* \rho U_t,f}\\
& = \frac{f(0)}{2}||U_t^*(i[ \rho ,A])U_t||^2_{U_t^* \rho U_t,f} = \frac{f(0)}{2}||i[ \rho ,A]||^2_{\rho ,f} = I^f_{\rho}(A).
\end{align*}

\begin{proposition}\label{disopra}
$$
\tilde{g} \leq \tilde{f} \quad \Longrightarrow 
I_{\rho}^f (A) \leq I_{\rho}^g (A) .
$$
\end{proposition}
\begin{proof}
Immediate consequence of Proposition \ref{boh}.
\end{proof}

Theorem 2 in \cite{Luo:2003a} is a particular case of the following result.

\begin{theorem} \label{main}
We have that for any $ f \in
{\cal F}_{op}^{\, r}$, for any $\rho \in {\cal D}^1_n$ and for any $A \in M_{n,sa}$
$$
I_{\rho}^f (A) \leq I_{\rho}^{SLD} (A) \leq \frac{1}{2f(0)}I_{\rho}^f (A).
$$
\end{theorem}
\begin{proof}
The first inequality is an immediate consequence of Proposition \ref{disopra}, Example \ref{serve} and Corollary \ref{basic}.
The second inequality is a consequence of Proposition \ref{bounds}, because we have
$$
|| \dot{\rho}_A ||_{\rho,f_{SLD}} \leq || \dot{\rho}_A ||_{\rho,f}
$$
and therefore
$$
\frac{f_{SLD}(0)}{2}|| \dot{\rho}_A ||^2_{\rho,f_{SLD}} \leq \frac{1}{4}|| \dot{\rho}_A ||^2_{\rho,f}
$$
so that
$$
 I_{\rho}^{f_{SLD}}(A)=
\frac{f_{SLD}(0)}{2}|| \dot{\rho}_A ||^2_{\rho,f_{SLD}}\leq
\frac{1}{2f(0)}\cdot\frac{f(0)}{2} \cdot || \dot{\rho}_A ||^2_{\rho,f}=\frac{1}{2f(0)}\cdot I_{\rho}^{f}(A).
$$

\end{proof}

A different proof can be given of the second inequality. It is more complicated but can shed light on Luo's proof and on the optimality of the constant $\frac{1}{2f(0)}$.

\begin{proposition} \label{alternative}
Let $k \geq 1$. The following inequalities are equivalent
$$\begin{array}{lrcl}
(i)& \qquad \qquad I_{\rho}^{SLD} (A) &\leq &k\cdot I_{\rho}^f (A) \qquad \qquad \forall A \in M_{n,sa}, \forall \rho \in {\cal D}^1_n, \\[12pt]
(ii)& \qquad \qquad m_{\tilde f}& \leq &\left(1-\frac{1}{k}\right)m_{{\cal A}}+\frac{1}{k}m_{\cal H}, \\[12pt]

(iii)& \qquad \qquad f(x) &\leq & 2kf(0)\cdot   \frac{1+x}{2}, \qquad \qquad \forall x>0.
\end{array}$$

\end{proposition}
\begin{proof}
Let $\left\{\varphi_i\right\}$ be a complete orthonormal base composed
of eigenvectors of $\rho$, and $\{ {\lambda}_i \}$ the corresponding
eigenvalues.
Set $a_{ij} \equiv \langle {A_0} {\varphi}_i |{\varphi}_j \rangle $. 
Note that $a_{ij} \not= A_{ij}:=$ the $i,j$ entry of $A$. 

As a consequence of the spectral theorem for commuting selfadjoint operators one gets the following formulas (see \cite{GibiliscoIsola:2007}):
$$
\Var_{\rho}(A) = {\rm Tr} ( \rho A_0^2) = \frac{1}{2}\sum_{i,j} ( \lambda _i + \lambda_j ) a_{ij} a_{ji}, 
$$
$$
{\cal C}^{\tilde f}_{\rho}(A_0)=\sum_{i,j} m_{\tilde f}(\lambda_i,\lambda_j)a_{ij}a_{ji}. 
$$

$(i) \Longleftrightarrow (ii)$.
\begin{align*}
k\cdot I_{\rho}^f (A) -I_{\rho}^{SLD} (A)
&=[k\cdot \Var_{\rho}(A)-k \cdot {\cal C}^{\tilde f}_{\rho}(A_0)]-[\Var_{\rho}(A)-{\cal C}^{\tilde f_{SLD}}_{\rho}(A_0)] \\
&= (k-1) \Var_{\rho}(A)+ {\cal C}^{\tilde f_{SLD}}_{\rho}(A_0)-k{\cal C}^{\tilde f}_{\rho}(A_0)\\
&=(k-1)\sum_{i,j} \frac{1}{2}\cdot(
\lambda _i + \lambda_j ) a_{ij} a_{ji}+\sum_{i,j} m_{\cal H}(\lambda_i,\lambda_j) a_{ij}a_{ji}-k\cdot \sum_{i,j}  m_{\tilde
f}(\lambda_i,\lambda_j) a_{ij}a_{ji} \\
&= k \sum_{i,j} \left[ \Bigl(1-\frac{1}{k} \Bigr) m_{\cal A}(\lambda _i,\lambda_j)+ \frac{1}{k} m_{\cal H}(\lambda_i,\lambda_j)- m_{\tilde
f}(\lambda_i,\lambda_j) \right] |a_{ij}|^2.  \\
\end{align*}
Therefore, because of the arbitrarity of both $\rho$ and  $A$, one has that
$$
kI_{\rho}^f (A) -I_{\rho}^{SLD} (A) \geq 0
$$
is equivalent to
$$
m_{\tilde f} \leq \Bigl(1-\frac{1}{k} \Bigr) m_{\cal A} + \frac{1}{k}m_{\cal H} .
$$

$(ii) \Longleftrightarrow (iii)$.
Suppose $x>0$, $x \not=1$. Then
$$
m_{\tilde f} \leq \Bigl(1-\frac{1}{k} \Bigr) m_{\cal A} + \frac{1}{k}m_{\cal H} 
$$
is equivalent to 
$$
{\tilde f}(x) \leq \Bigl( 1-\frac{1}{k} \Bigr)\left(\frac{1+x}{2} \right)  + \frac{1}{k} \left( \frac{2x}{x+1}\right) \qquad \qquad \forall x >0
$$

which, using the definition of $\tilde f$, can be transformed into
$$
2kf(0)\cdot  \frac{1+x}{2} \geq f(x) \qquad \qquad \forall x >0
$$
and this ends the proof.
\end{proof}

\begin{example}
In the case of the Wigner-Yanase metric one has $f_{WY}(0)=\frac{1}{4}$ and ${\tilde f}_{WY}(x)=\sqrt{x}$. The inequality of Proposition \ref{alternative}(ii) (when $k=2=\frac{1}{2f_{WY}(0)}$) states that
$$
m_{\cal G} \leq \frac{1}{2}(m_{\cal A}+m_{\cal H})
$$
that is the geometric mean is smaller then the ``midpoint" between arithmetic and harmonic mean. The calculations used by Luo in the proof of inequality (1.1) can be seen as an application of the above inequality.
\end{example}

We now prove that $\frac{1}{2f(0)}$ is the best constant we can have in Theorem \ref{main}.

\begin{proposition} Let $1 \leq k \leq \frac{1}{2f(0)}$. The inequality
$$
I_{\rho}^{SLD} (A) \leq k\cdot I_{\rho}^f (A) \qquad \qquad \forall A \in M_{n,sa}, \forall \rho \in {\cal D}^1_n 
$$
is false.
\end{proposition}
\begin{proof}
From the hypothesis we get that the inequality
$$
f(x) \leq 2kf(0)\cdot  \frac{1+x}{2} \qquad \qquad \forall x>0 
$$
cannot be true, otherwise one would have
$$
1=f(1) \leq 2kf(0) <1
$$
which is absurd. From Proposition \ref{alternative} we get the conclusion.
\end{proof}

\section{The inequality on the Bloch sphere}

As an example we discuss in detail what happens for $2\times2$ matrices. We show that also in this case  the constant $\frac{1}{2f(0)}$ is optimal. 
The final Example in \cite{Luo:2003a} is a particular case of this discussion.

Recall that the Pauli matrices are the following
$$
\sigma_1= \begin{pmatrix}
0 & 1  \cr
1 & 0  \cr
\end{pmatrix},
\qquad
\sigma_2=\begin{pmatrix}
0 & -i  \cr
i & 0  \cr
\end{pmatrix},
\qquad
\sigma_3=\begin{pmatrix}
1 & 0  \cr
0 & -1  \cr
\end{pmatrix}.
$$

A generic $2 \times 2$ density matrix in the Stokes parameterization is written as
$$
\rho= \frac{1}{2}\begin{pmatrix}
1+x & y+iz  \cr
y-iz & 1-x  \cr
\end{pmatrix}= \frac{1}{2}(I + x\sigma_1+y\sigma_2+z\sigma_3),
$$
where $(x,y,z) \in {\mathbb R}^3$,  and $x^2+y^2+z^2 \leq 1$.
Let $r := \sqrt{x^2+y^2+z^2}\in[0,1]$. The eigenvalues of $\rho$ are $\lambda_1=\frac{1-r}{2}$ and $\lambda_2=\frac{1+r}{2}$.

\begin{proposition}
$$
I_{\rho}^f(A)=\left[1-m_{\tilde f}(1-r,1+r) \right] \cdot |a_{12}|^2.
$$
\end{proposition}
\begin{proof}
We use notation as in the proof of Proposition \ref{alternative}. Observe that
$$
\frac{\lambda_i+\lambda_j}{2}-m_{\tilde f}(\lambda_i,\lambda_j) =
\begin{cases}
	0, & i=j,\\
	\frac{1}{2}-m_{\tilde f}(\lambda_i,\lambda_j), & i\neq j.
\end{cases}
$$
Therefore
\begin{align*}
I_{\rho}^f(A)
&=\sum_{i,j}\left[\frac{\lambda_i+\lambda_j}{2}-m_{\tilde f}(\lambda_i,\lambda_j)\right]\cdot|a_{ij}|^2\\
&=\left[\frac{1}{2}-m_{\tilde f}(\frac{1-r}{2},\frac{1+r}{2})\right]|a_{12}|^2+\left[\frac{1}{2}-m_{\tilde f}(\frac{1+r}{2},\frac{1-r}{2})\right]|a_{21}|^2\\
&=\left[1-m_{\tilde f}(1-r,1+r) \right] \cdot |a_{12}|^2.
\end{align*}
\end{proof}

\begin{corollary} If $r\not=0$ then
$$
I_{\rho}^{SLD}(A)=\left[ \frac{r^2}{1-m_{\tilde f}(1-r,1+r)}\right]\cdot I_{\rho}^f(A).
$$
\end{corollary}
\begin{proof}
If $f_{SLD}(x)=\frac{1+x}{2}$ then ${\tilde f}_{SLD}=\frac{2x}{x+1}$. In this case
$$
m_{{\tilde f}_{SLD}}(1-r,1+r)= (1+r){\tilde f_{SLD}}\left(\frac{1-r}{1+r}\right) =1-r^2.
$$
Therefore, from the above proposition
$$
I_{\rho}^{SLD}(A)=\left[ 1-m_{{\tilde f}_{SLD}}(1-r,1+r) \right] \cdot |a_{12}|^2=\left[ 1-(1-r^2) \right] \cdot |a_{12}|^2=r^2\cdot |a_{12}|^2
$$
and this ends the proof.
\end{proof}

\begin{example}
In the case $f_{WY}(x)=\left( \frac{1+\sqrt{x}}{2} \right)^2$ one has ${\tilde f}_{WY} (x)=\sqrt{x}$. In this case (see \cite{Luo:2003a})
$$
I_{\rho}^{SLD}(A)=\left[ \frac{r^2}{1-m_{\tilde f_{WY}}(1-r,1+r)}\right]\cdot I_{\rho}^{WY}(A)=\left[ \frac{r^2}{1-\sqrt{1-r^2}}\right]\cdot I_{\rho}^{WY}(A)
=[ 1+\sqrt{1-r^2} ] \cdot I_{\rho}^{WY}(A).
$$
\end{example}

\begin{remark}
\end{remark}
Note that for any regular $f$  the function  $\tilde f$ is not regular and therefore
$$
\lim_{r \to 1} \frac{r^2}{1-m_{\tilde f}(1-r,1+r)} = \lim_{r \to 1} \frac{r^2}{1-(1+r){\tilde f}\left(\frac{1-r}{1+r}\right)}
=\frac{1}{1-{\tilde f}(0)}=1.
$$
We already know such a result because the case $r=1$ is that of pure states where all the $f$-informations coincide with variance.

\begin{proposition}
If $f$ is regular then
$$
\lim_{r \to 0} \frac{r^2}{1-m_{\tilde f}(1-r,1+r)} = - \frac{1}{2{\tilde f}''(1)}=\frac{1}{2f(0)}.
$$
\end{proposition}
\begin{proof}
Let $g(r) := 1-m_{\tilde f}(1-r,1+r)$. For any $f \in {\cal F}_{op}$ one has $f'(1)=\frac{1}{2}$ (because of symmetry) and this implies that 
 $g(0)=g'(0)=0$.  Therefore we have to use twice the De L'Hopital theorem. 
 An easy calculation shows that ${\tilde f}''(1)=-f(0)$, therefore we get
\begin{align*}
\lim_{r \to 0} \frac{r^2}{1-m_{\tilde f}(1-r,1+r)} 
&=\lim_{r \to 0} \frac{\frac{{\rm d}^2}{{\rm d}r^2}r^2}{\frac{{\rm d}^2}{{\rm d}r^2}\left[1-m_{\tilde f}(1-r,1+r)\right]} =\lim_{r \to 0} \frac{2}{-\frac{4}{(1+r)^3}{\tilde f}''\left(\frac{1-r}{1+r}\right)}\\
&=\frac{2}{-4{\tilde f}''(1)} 
= \frac{1}{2f(0)}.
\end{align*}

\end{proof}

From the above Proposition we get a different proof of the fact that the constant $\frac{1}{2f(0)}$ is optimal also in the $2 \times 2$ matrix case.

\end{document}